# A Novel Approach for Single Gene Selection Using Clustering and Dimensionality Reduction

E.N.Sathishkumar, K.Thangavel, T.Chandrasekhar

**Abstract**—We extend the standard rough set-based approach to deal with huge amounts of numeric attributes versus small amount of available objects. Here, a novel approach of clustering along with dimensionality reduction; Hybrid Fuzzy C Means-Quick Reduct (FCMQR) algorithm is proposed for single gene selection. Gene selection is a process to select genes which are more informative. It is one of the important steps in knowledge discovery. The problem is that all genes are not important in gene expression data. Some of the genes may be redundant, and others may be irrelevant and noisy. In this study, the entire dataset is divided in proper grouping of similar genes by applying Fuzzy C Means (FCM) algorithm. A high class discriminated genes has been selected based on their degree of dependence by applying Quick Reduct algorithm based on Rough Set Theory to all the resultant clusters. Average Correlation Value (ACV) is calculated for the high class discriminated genes. The clusters which have the ACV value as 1 is determined as significant clusters, whose classification accuracy will be equal or high when comparing to the accuracy of the entire dataset. The proposed algorithm is evaluated using WEKA classifiers and compared. Finally, experimental results related to the leukemia cancer data confirm that our approach is quite promising, though it surely requires further research.

—————————— ◆ ——————————

## 1 INTRODUCTION

THE DNA microarray technology provides enormous quantities of biological information about genetically conditioned susceptibility to diseases. The data sets acquired from microarrays refer to genes via their expression levels. Microarray production starts with preparing two samples of mRNA, as illustrated by Fig. 1. The sample of interest is paired with a healthy control sample. The fluorescent red/green labels are applied to both samples. The procedure of samples mixing is repeated for each of thousands of genes on the slide. Fluorescence of red/green colors indicates to what extent the genes are expressed. The gene expressions can be then stored in numeric attributes, coupled with, e.g., clinical information about the patients. Given thousands of genes-attributes against hundreds of objects, we face a "few-objects-many-attributes" problem [19]. Dimensionality reduction in gene expression data can be critical for a number of reasons. First, for large number of genes or feature set, the processing of all available genes may be computationally infeasible. Second, many of the available features may be redundant and noise-dominated or irrelevant to the classification task at hand. Third, high-dimensionality is also a problem if the number of variables is much larger than the number of data points available. In such a scenario, dimensionality reduction is crucial in order to overcome the curse of dimensionality [7],[11] and allow for meaningful data analysis. For the above reasons, feature selection is important for gene expression data analysis.

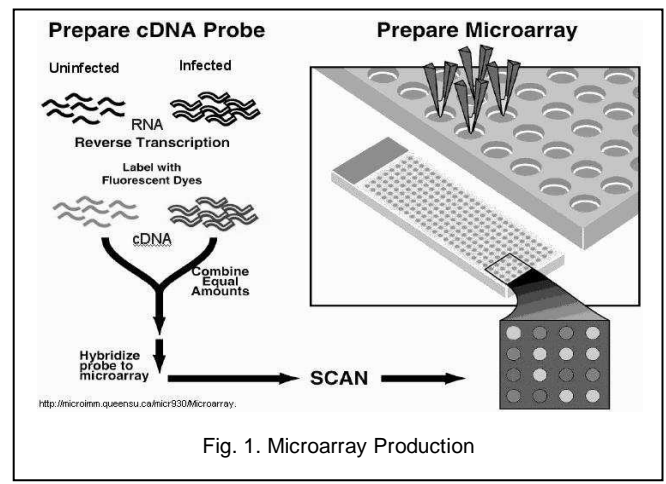

Fig. 1. Microarray Production

A problem with gene expression analysis is often the selection of significant variables (feature selection) within the data set that would enable accurate classification of the data to some output classes. These variables may be potential diagnostic markers too. There are good reasons for reducing the large number of variables: First, an opportunity to scrutinize individual genes for further medical treatment and drug development. Second, dimension reduction to reduce the computational cost. Third, reducing the number of redundant and unnecessary variables can improve inference and classification. Fourth, more interpretable features or characteristics that can help identify and monitor the target diseases or functions types.

In this paper, gene or features will be selected from a group, such that the genes in a group will be similar. Gene clustering identifies groups of genes that exhibit similar expression profiles across samples. Clustering is a widely used technique for analysis of gene expression data. Most clustering methods

————————————————

- E.N. Sathishkumar is currently pursuing Ph.D., in Computer Science in Periyar University, Salem, India. E-mail: en.sathishkumar@yahoo.in
- K. Thangavel is currently working as Professor and Head, Department of Computer Science in Periyar University, Salem, India. E-mail: drktvelu@yahoo.com
- T. Chandrasekhar is currently pursuing Ph.D., in Computer Science in Bharthiyar University, India. E-mail: ch_ansekh80@rediffmail.com

group genes based on the distances, while few methods group according to the similarities of the distributions of the gene expression levels. Clustering is the process of finding groups of objects such that the objects in a group will be similar (or related) to one another and different from (or unrelated to) the objects in other groups. A good clustering method will produce high quality clusters with high intra-cluster similarity and low inter-cluster similarity. The quality of a clustering result depends on both the similarity measure used by the method and its implementation and also by its ability to discover some and all of the hidden patterns [16].

Feature Selection algorithm aims at finding out a subset of the most representative features according to some objective function in discrete space. The algorithms of FS are always greedy. Our feature selection will be based on rough set; The Rough set approach to feature selection consists in selecting a subset of features which can predict the classes as well as the original set of features. The optimal criterion for Rough set feature selection is to find shortest or minimal reducts while obtaining high quality classifiers based on the selected features. In this paper, we introduce a novel method of using Fuzzy C-Means clustering along with rough set attribute reduction (Quick Reduct) for single gene selection. The attribute selection method is perform for all clusters which are obtained by the applying FCM algorithm.

This paper is organized as follows. The next section describes about various methods and in section 3 briefs the proposed gene selection algorithm. In section 4, experimental results are listed. The discussions of these results are given. Section 5 briefs about WEKA classification and its classification results. Finally, this paper is concluded in section 6.

## 2 METHODS
### 2.1 Fuzzy C-Means Clustering

Fuzzy clustering allows each feature vector to belong to more than one cluster with different membership degrees (between 0 and 1) and vague or fuzzy boundaries between clusters. Fuzzy C-Means (FCM) is a method of clustering which allows one piece of data to belong to two or more clusters[2],[14]. This method (developed by Dunn in 1973 and improved by Bezdek in 1981) is frequently used in pattern recognition. The Fuzzy C Means algorithm is given below:

**Algorithm 1**: Fuzzy C-Means clustering algorithm [14]

Step-1: Randomly initialize the membership matrix using this equation,

$$\sum_{j=1}^{C} \mu_j(x_i) = 1 \qquad i = 1, 2 \ldots k$$

Step-2: Calculate the Centroid using equation,

$$Cj = \frac{\sum_i [\mu_j(x_i)]^m x_i}{\sum_i [\mu_j(x_i)]^m}$$

Step-3: Calculate dissimilarly between the data points and Centroid using the Euclidean distance.

$$D_i = \sqrt{(x_2 - x_1)^2 + (y_2 - y_1)^2}$$

Step-4: Update the New membership matrix using the equation,

$$\mu_j(x_i) = \frac{[\frac{1}{d_{ji}}]^{1/m-1}}{\sum_{k=1}^{c}[\frac{1}{d_{ki}}]^{1/m-1}}$$

Here **m** is a fuzzification parameter.
The range **m** is always [1.25, 2]

Step-5: Go back to Step 2, unless the centroids are not changing.

### 2.2 K-Means Discretization

Many data mining techniques often require that the attributes of the data sets are discrete. Given that most of the experimental data are continuous, not discrete, the discretization of the continuous attributes is an important issue. The goal of discretization is to reduce the number of possible values a continuous attribute takes by partitioning them into a number of intervals. K-Means discretization method is used in this paper, in gene expression data set each gene attribute are clustered with K-Means, replaces the attribute values with the cluster membership labels. These labels will be act as discrete values for gene expression data set.

### 2.3 Quick Reduct Algorithm

Rough set theory (Pawlak, 1991) is a formal mathematical tool that can be applied to reducing the dimensionality of dataset. The rough set attribute reduction method removes redundant input attributes from dataset of discrete values, all the while making sure that no information is lost. The reduction of attributes is achieved by comparing equivalence relations generated by sets of attributes. Attributes are removed so that the reduced set provides the same quality of classification as the original. A reduct is defined as a subset R of the conditional attribute set C such that $\gamma_R(D) = \gamma_C(D)$. A given dataset may have many attribute reduct sets, so the set R of all reducts is defined as:

$$R_{all} = \{X \mid X \subseteq C, \gamma_X(D) = \gamma_C(D);$$

$$\gamma_{X-\{a\}}(D) \neq \gamma_X(D)), \forall a \in X\}. \qquad (1)$$

The intersection of all the sets in $R_{all}$ is called the core, the elements of which are those attributes that cannot be eliminated without introducing more contradictions to the representation of the dataset. For many tasks (for example, feature selection), a reduct of minimal cardinality is ideally searched for a single element of the reduct set $R_{min} \subseteq R_{all}$:

$$R_{min} = \{X \mid X \in R_{all}, \forall Y \in R_{all}, |X| \leq |Y|\}. \qquad (2)$$

The Quick Reduct algorithm shown below, it searches for a minimal subset without exhaustively generating all possible subsets. The search begins with an empty subset; attributes which result in the greatest increase in the rough set dependency value that is added iteratively. This process continues until the search produces its maximum possible dependency value for that data set $\gamma_C(D)$. Note that this type of search does not guarantee a minimal subset and may only discover a local minimum. Such techniques may be found in [1],[7],[8],[14],[17]

---

**Algorithm 2:** Quick Reduct (C, D)

---

C, the set of all conditional features;
D, the set of decision features.
(a) R ← {}
(b) **Do**
(c) T ← R
(d)     ∀ x ∈(C-R)
(e)     **if** $\gamma_{R\cup\{x\}}(D) > \gamma_T(D)$

Where $\gamma_R(D) = \frac{card\ (POS_R(D))}{card\ (U)}$

(f)     T ← R∪{x}
(g) R ← T
(h) **until** $\gamma_R(D) == \gamma_C(D)$
(i) **return** R

---

## 2.4 Average Correlation Value

Average Correlation Value is used to evaluate the homogeneity of a cluster. Matrix A= ($A_{ij}$) has the ACV which is defined by the following function,

$$ACV\ (A) = max\left\{\frac{\sum_{i=1}^{m}\sum_{j=1}^{m}|C\_row_{ij}|-m}{m^2-m}, \frac{\sum_{p=1}^{n}\sum_{q=1}^{n}|C\_col_{pq}|-n}{n^2-n}\right\} \quad (3)$$

Where C_row$_{ij}$ – is the correlation coefficient between rows i and j, C_col$_{pq}$ is the correlation coefficient between columns p and q, ACV approaching 1 denote a significant cluster. Such technique may be found in [12].

## 3 HYBRID FUZZY C-MEAN-QUICKREDUCT (FCMQR) ALGORITHM

The proposed FCMQR algorithm logically consists of three steps: (i) grouping the similar genes, (ii) feature selection from group, (iii) finding ACV and selecting representative features. The purpose of the algorithm is to select a subset of features R = {$RG_1$, $RG_2$,…,$RG_r$} from the original gene set G = {$G_1$, $G_2$,…,$G_n$} where n is the dimension of gene feature vectors and r<n is the number of selected features that having ACV =1. A feature $G_{best}$ is included in the subset R, if for this $G_{best}$, the subset R gives the highest classification accuracy. The algorithm of FCMQR method is described as follows.

---

**Algorithm 3:** Hybrid Fuzzy C-Mean-QuickReduct (FCMQR)

---

**Inputs:** Gene expression data contains *n* genes and a *m* class variable, G = {$G_1$, $G_2$… $G_n$} and D= {$D_1$, $D_2$… $D_m$}
**Output:** $G_{best}$ – Selected gene
Step 1: set *k*=5 and $G_{best}$ ← {}
Step 2: Do, Gene wise cluster using FCM
    i) Random membership initialization

$$\sum_{j=1}^{C}\mu_j(x_i) = 1$$

ii) Calculate the Centroid $Cj = \frac{\sum_i [\mu_j(x_i)]^m x_i}{\sum_i [\mu_j(x_i)]^m}$

iii) Calculate dissimilarly between the data points and Centroid using the Euclidean distance.
iv) Update new membership matrix

$$\mu_j(x_i) = \frac{[\frac{1}{d_{ji}}]^{1/m-1}}{\sum_{k=1}^{c}[\frac{1}{d_{ki}}]^{1/m-1}}$$

v) Go back to ii, unless the centroids are not changing.

Step 3: Get $GC_i$= {$g_1$, $g_2$… $g_q$} from Step 2.
Step 4: Discretize the $GC_i$ by applying K-means Discretization
Step 5: for i= 1 to *k*
        Do, Quick Reduct($GC_i$,D) to select
            $RC_i$ = {$R_1$, $R_2$…$R_r$} according to Alg.2
                where $RC_i$ ⊆ $GC_i$
        End
    End
Step 6: Compute ACV for all refined $RC_i$ according to Eqs.3
Step 7: Collect all the genes from clusters, where ACV=1
        $R_k$ = {$RC_i$ / ACV ($RC_i$) = 1} = {$RG_1$, $RG_2$,…,$RG_r$}
            where r = no. of genes in acv=1 clusters
Step 8: Repeat step 2 to 6, for *k* = 7 and etc.
                where *k*= no. of clusters we need.
Step 9: Let $G_{best}$ = $\bigcap_{k\epsilon Rk} R_k$
Step 10: Return $G_{best}$

---

## 4 EXPERIMENTAL RESULTS

### 4.1 Data Set

We use leukemia data set which is available in the website: http://datam.i2r.a-star.edu.sg/ datasets/krbd/ [15]. Our initial leukemia data set consisted of 38 bone marrow samples (27 ALL, 11 AML) obtained from acute leukemia patients at the time of diagnosis. RNA prepared from bone marrow mononuclear cells was hybridized to high-density oligonucleotide microarrays, produced by Affymetrix and containing probes over 7129 from 6817 human genes [18].

### 4.2 Cluster Analysis and Gene Selection

Gene clustering identifies group of genes that exhibit similar expression profiles across samples. Fuzzy C-Means clustering is used to cluster the similar characteristics of genes $GC_i$. Before clustering, need to specify the number of clusters. The optimal number of clusters is difficult to determine, because it may depend on different sets of genes under investigation. In this study, the number of clusters is chosen to be five and seven (*k*=5, 7), then leukemia data set will divide *k* number of groups using Fuzzy C-Means clustering techniques. After clustering, features or genes will be selected from a similar gene cluster $GC_i$. Rough Quick Reduct has been used as feature selection method. The data studied by rough set are mainly organized in the form of decision tables. One decision table can be represented as S = (U, A=$GC_i$ U D), where U is the set of samples in cluster $GC_i$ (i=1 to *k*), $GC_i$ the condition attribute set and D the decision attribute set. We can represent every gene expression data with the decision table like Table 1.

TABLE 1

ROUGH SET DATA DECISION TABLE

| Samples | Cluster $GC_i$ (Condition attributes) | | | | Decision attributes |
|---|---|---|---|---|---|
| | Gene1 | Gene 2 | … | Gene n | Class label |
| 1 | g(1,1) | g(1,2) | … | g(1,n) | Class(1) |
| 2 | g(2,1) | g(2,2) | … | g(2,n) | Class(2) |
| … | … | … | … | … | … |
| p | g(m,1) | g(m,2) | … | g(m,n) | Class(m) |

In the decision table, there are *m* samples and *n* genes in cluster $GC_i$. Every sample is assigned to one class label. Each gene is a condition attribute and each class is a decision attribute. *g(m, n)* signifies the expression level of gene *n* in sample *m* [15]. Before applying feature selection algorithm all the conditional attributes (samples) are discretized using K-Means discretization. After feature selection, to evaluate the Average Correlation Value for selected genes from each cluster. ACV approaching 1 denote a significant cluster and it is evaluating the homogeneity of a cluster. Table 2 and 3 shows the selected genes from particular cluster by applying Quick Reduct and Average Correlation Value for that genes.

TABLE 2
SELECTED GENES WHEN CLUSTER K=5

| Cluster | All Genes ($GC_i$) | QR Selected Genes ($RC_i$) | ACV |
|---|---|---|---|
| Cluster 1 | 203 | #1962, #2288 | 0.6011 |
| *Cluster 2* | 1453 | #154, #3252 | 1 |
| Cluster 3 | 75 | #42,#45, #1707 | 0.6266 |
| *Cluster 4* | 42 | #930, #1765, #5711 | 1 |
| Cluster 5 | 5356 | #43,#1962,#6361 | 0.6266 |

Table 2, depict the similar expression genes when *k*=5, and shows selected genes ($RC_i$) after applied Quick Reduct. Based on Average Correlation Values, we determine cluster 2 and 4 are significant clusters $R_k$. In that cluster genes having high classification accuracy compare to other genes.

TABLE 3
SELECTED GENES WHEN CLUSTER K=7

| Cluster | All Genes ($GC_i$) | QR Selected Genes ($RC_i$) | ACV |
|---|---|---|---|
| Cluster 1 | 766 | #930,#5350,#6104, #6670 | 0.6559 |
| *Cluster 2* | 5003 | #29, #1884 | 1 |
| Cluster 3 | 74 | #1775, #4116, #6193 | 0.6280 |
| *Cluster 4* | 1007 | #79, #3252 | 1 |
| Cluster 5 | 38 | #44, #47, #896, #445 | 0.5079 |
| Cluster 6 | 52 | #42, #45, #67 | 0.6188 |
| *Cluster 7* | 189 | #1674, #2288 | 1 |

Table 3, shows similar expression genes when *k*=7 and depict selected genes ($RC_i$) after Quick Reduct. Based on Average Correlation Values, we determine Cluster 2, 4 and 7 are significant clusters.

## 5 WEKA CLASSIFICATION

The classifier tool WEKA [1],[11],[13] is open source java based machine-learning. It brings together many machine learning algorithm and tools under a common frame work. The WEKA is a well known package of data mining tools which provides a variety of known, well maintained classifying algorithms. This allows us to do experiments with several kinds of classifiers quickly and easily. The tool is used to perform benchmark experiment. Some of the classifiers we used in our experiment are bayes.NaiveBayes, trees.J48, rules.Decision Table and lazy.K-Star.

TABLE 4
CLASSIFICATION ACCURACY OF GENES WHEN K=5

| Classifier | Entire data 7129 genes (G) | QR selected 13 genes | ACV=1 5 genes ($R_k$) |
|---|---|---|---|
| Naïve | 94.1176 | 97.0588 | 97.8805 |
| D. Table | 88.2353 | 94.1176 | 94.1176 |
| J48 | 91.1765 | 97.0588 | 97.0588 |
| K-Star | 58.8235 | 94.1176 | 95.6568 |

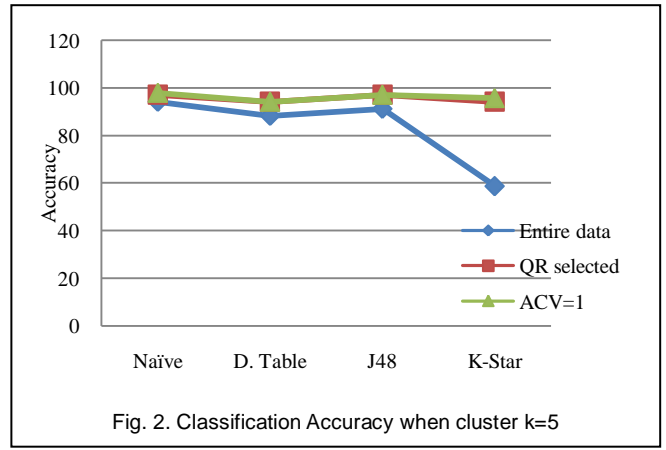

Fig. 2. Classification Accuracy when cluster k=5

Fig. 2, Denotes classification accuracy for leukemia data when cluster *k*=5. We obtained thirteen genes from entire genes by applying Quick Reduct. Out of the thirteen genes which genes having ACV=1 that five genes are selected as best $R_k$. The classification accuracy for those five; #154, #3252, #930, #1765 and #5711 genes is equal or high when compared to entire genes (7129) and Quick Reduct selected genes.

TABLE 5
CLASSIFICATION ACCURACY OF GENES WHEN K=7

| Classifier | Entire data 7129 genes (G) | QR selected 20 genes | ACV=1 6 genes ($R_k$) |
|---|---|---|---|
| Naïve | 94.1176 | 97.0588 | 97.0588 |
| D. Table | 88.2353 | 91.1765 | 94.1176 |
| J48 | 91.1765 | 97.0588 | 91.1765 |
| K-Star | 58.8235 | 94.1176 | 95.6568 |

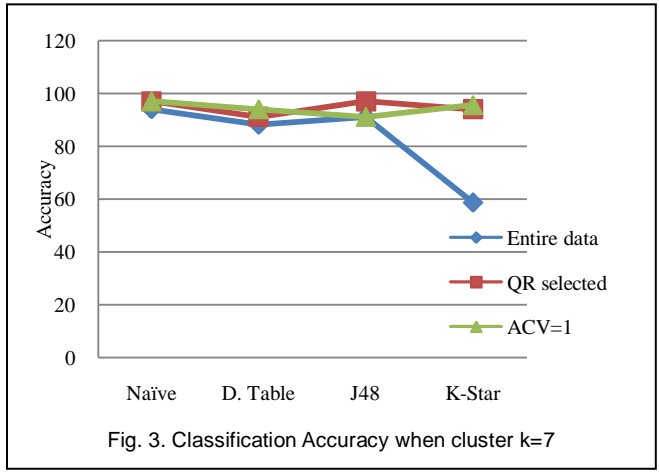

Fig. 3. Classification Accuracy when cluster k=7

Fig. 3, Denotes classification accuracy of leukemia data when $k$=7. We obtained twenty genes from entire genes by applying QuickReduct. Out of the twenty genes which genes having ACV=1 that six genes are selected as high class discriminated genes $R_k$. The classification accuracy for those six; #29, #1884, #79, #3252, #1674 and #2288 genes is equal or high when compared to entire genes and QuickReduct selected genes.

TABLE 6
MARKER GENE SELECTED FROM SIGNIFICANT CLUSTERS GENES

| Cluster | $R_k$ Genes | Selected gene |
|---|---|---|
| Cluster $k$=5 | #154, #3252, #930, #1765, #5711 | #3252 ($G_{best}$) |
| Cluster $k$=7 | #29, #1884, #79, #3252, #1674, #2288 | |

Table 6 shows, In the leukemia dataset, when cluster $k$=5, gene #154, #3252, #930, #1765 and #5711 are identified; when $k$=7, gene #29, #1884, #3252, #79, #1674 and #2288 are identified. Among the significant clusters (ACV=1) genes have the classification accuracy higher than 88.2353%. Finally, we get $G_{best} = \bigcap_{k \in Rk} R_k$ is gene #3252 has 97.0588% accuracy and which is common to all experiment. We denote the expression level of gene x by g(x). Two decision rules induced by gene #3252 are:

If g (#3252) > 643, then AML; If g (#3252) ≤ 643, then ALL.

### 5.1 Comparison of Classification Results

The leukemia dataset has been well studied by many researchers [3], [4], [5], [6], [8], [9], [10]. Although there are a few reports on the use of a single gene to distinguish the AML from the ALL, a majority of investigators conduct the classification with more than one gene, even tens or hundreds. In[15], the authors present the classification outcomes of 31 out of 34 samples correctly classified with one common gene (#4847): X.Wang & O.Gotoh, we correctly classify 33 out of 34 samples using a selected gene (#3252). Classification accuracy for existing and proposed method selected single genes shown in table 7.

TABLE 7
COMPARISON BETWEEN EXISTING AND PROPOSED METHOD

| Classifier | Existing Method gene #4847 | Proposed FCMQR gene #3252 |
|---|---|---|
| Naïve | 91.1765 | 97.0588 |
| D.Table | 85.2941 | 91.1765 |
| J48 | 88.2353 | 91.1765 |
| K-Star | 91.1765 | 97.0588 |

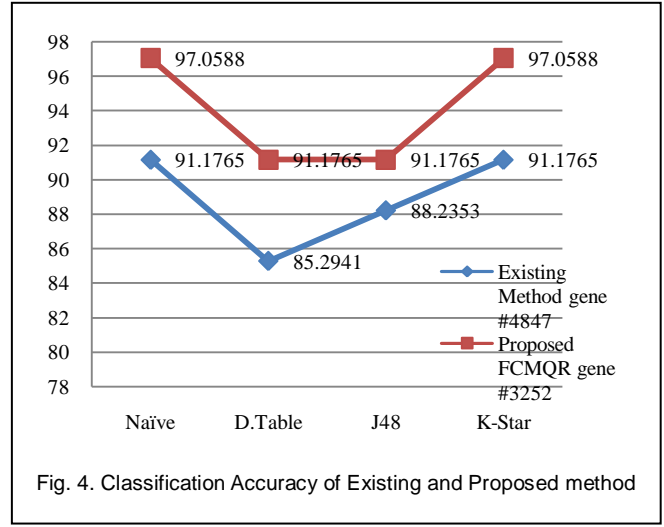

Fig. 4. Classification Accuracy of Existing and Proposed method

Regarding leukemia datasets, the best classification results reported in our and existing works are shown in Table 7 and Fig 4, respectively. These tables demonstrate that our FCMQR algorithm perform comparatively well in leukemia dataset.

## 6 CONCLUSION

The work has been proposed for improving the gene selection method in a simple and efficient way. Here a novel approach combining clustering and rough set attribute reduction method FCMQR has been proposed for gene expression data. Informative genes are selected using their classification accuracy. Fuzzy C-Means clustering, Rough Quick Reduct and Average Correlation Value methods are studied and implemented successfully for gene selection. The proposed work gives sparse and interpretable classification accuracy, compared to other gene selection method on leukemia gene expression data set. The classification accuracy of genes has been done using WEKA classifier.

### ACKNOWLEDGMENT

The present work is supported by Special Assistance Programme of University Grants Commission, New Delhi, India (Grant No. F.3-50/2011(SAP-II).

The first author immensely acknowledges the partial financial assistance under University Research Fellowship, Periyar University, Salem, Tamilnadu, India.

## REFERENCES

[1] C. Velayutham and K. Thangavel, "Unsupervised Quick Reduct Algorithm using Rough Set Theory", Journal of Electronic Science and Technology, vol. 9, no. 3, Sep 2011.

[2] Thangavel K, Velayutham C,"Mammogram Image Analysis: Bio-Inspired Computational Approach "Proceedings of the International Conf. on SocProS 2011, AISC 131, pp. 883–892.

[3] Furey, T.-S., Cristianini, N., Duffy, N., Bednarski, D.-W., Schummer, M., and Haussler, D., Support vector machine classification and validation of cancer tissue samples using microarray expression data, Bioinformatics, 16(10):906-914, 2000.

[4] Geman, D., d'Avignon, C., Naiman, D.-Q., and Winslow, R.-L., Classifying gene expression profiles from pairwise mRNA comparisons, Stat Appl Genet Mol Biol 3:Article19, 2004.

[5] Golub, T.-R., Slonim, D.-K., Tamayo, P., Huard, C., Gaasenbeek, M., Mesirov J.P., Coller, H., Loh M.-L., Downing, J.-R., Caligiuri, M.-A., et al., Molecular classification of cancer: class discovery and class prediction by gene expression monitoring, Science, 286(5439):531-537, 1999.

[6] Lamba, J.-K., Pounds, S., Cao, X., Downing J.-R., Campana, D., Ribeiro, R.-C., Pui, C.H., and Rubnitz, J.-E., Coding polymorphisms in CD33 and response to gemtuzumab ozogamicin in pediatric patients with AML: a pilot study. Leukemia, 23(2):402-404, 2009.

[7] Lijun Sun, Duoqian Miao and Hongyun Zhang," Gene Selection with Rough Sets for Cancer Classification", Fourth International Conference on Fuzzy Systems and Knowledge Discovery (FSKD 2007), IEEE: 0-7695-2874-0/07.

[8] Li, D., and Zhang, W., Gene selection using rough set theory, Proc. 1st International Conference on Rough Sets and Knowledge Technology, 778–785, 2006.

[9] Li, J., and Wong, L., Identifying good diagnostic gene groups from gene expression profiles using the concept of emerging patterns, Bioinformatics, 18(5):725-734, 2002.

[10] Momin, B.-F., and Mitra, S., Reduct generation and classification of gene expression data, Proc. 1st International Conference on Hybrid Information Technology, 699-708, 2006. Sun, L., Miao, D., and Zhang, H., Efficient gene selection with rough sets from gene expression data, Proc. 3rd International Conference on Rough Sets and Knowledge Technology, 164–171, 2008.

[11] M. Dash and H. Liu, "Feature selection for classification", Intelligent Data Analysis, vol. 1, no. 3, pp. 131–156, 1997.

[12] R. Rathipriya, Dr. K.Thangavel and J.Bagyamani, "Evolutionary Biclustering of Clickstream Data", International Journal of Computer Science Issues, Vol. 8, Issue 3, No 1, May 2011. ISSN (Online): 1694-0814.

[13] Tan, A.-C., and Gilbert, D., Ensemble machine learning on gene expression data for cancer classification, Appl Bioinformatics, 2(3 Suppl):S75-83, 2003.

[14] T. Chandrasekhar, K. Thangavel and E.N. Sathishkumar, "Verdict Accuracy of Quick Reduct Algorithm using Clustering and Classification Techniques for Gene Expression Data", International Journal of Computer Science Issues, Vol. 9, Issue 1, No 1, January 2012. ISSN (Online): 1694-0814.

[15] Xiaosheng Wang and Osamu Gotoh, "Cancer Classification Using Single Genes", pp 179-188, Available: www.jsbi.org /pdfs/journal1 /GIW09/GIW09017.pdf

[16] Xu R. and Wunsch D., 2005. "Survey of clustering algorithms", IEEE Trans. Neural Networks, Vol. 16, No. 3, pp. 645-678.

[17] Z.Pawlak, Rough Sets: Theoretical Aspects of Reasoning about Data, Dordrecht: Kluwer Academic Publishers, 1991.

[18] T.R. Golub et.al, "Molecular Classification of cancer: Class Discovery and Class Prediction by Gene Expression Monitoring", www.sciencemag.org, vol 286, Oct.1999

[19] Dominik Slezak, "Rough Sets and Few-Objects-Many-Attributes Problem: The Case Study of Analysis of Gene Expression Data Sets", Frontiers in the Convergence of Bioscience and Information Technologies, IEEE 2007, pp. 437-440.

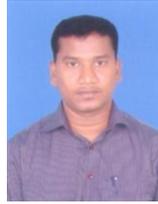

**E.N.Sathishkumar** was born in Elachipalayam at 1986, Tamilnadu, India. He received his Master of Science in Information Technology from Anna University, Coimbatore in 2009. He obtained his Master of Philosophy form the Department of Computer Science, Periyar University, Salem, India in 2011. He is pursuing his Ph.D., in Computer Science at Periyar University under the guidance of Dr. K.Thangavel.
His area of interests includes Data Mining, Rough Set, Bioinformatics and Neural Network.

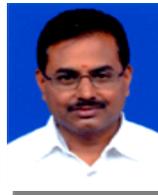

**Dr. K.Thangavel** was born in Namakkal at 1964, Tamilnadu, India. He received his Master of Science from the Department of Mathematics, Bharathidasan University in 1986, and Master of Computer Applications Degree from Madurai Kamaraj University, India in 2001. He obtained his Ph.D. Degree from the Department of Mathematics, Gandhigram Rural Institute- Deemed University, Gandhigram, India in 1999.
Currently he is working as Professor and Head, Department of Computer Science, Periyar University, Salem. He is a recipient of Tamilnadu Scientist award for the year 2009. His area of interests includes Medical Image Processing, Bioinformatics, Artificial Intelligence, Neural Network, Fuzzy logic, Data Mining and Rough Set.

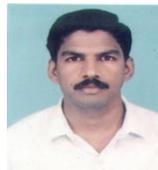

**T.Chandrasekhar** was born in Karur at 1980, Tamilnadu, India. He is received the Master of Science in information technology and management in 2003 and his M.Phil (Computer Science) Degree in 2004, from Bharathidasan University, Trichy, India. He is pursuing his Ph.D in Bharathiar University in Computer Science under the guidance of Dr. K.Thangavel.
Currently he is working as Guest lecturer, Department of Computer Science, Periyar University, Salem, Tamilnadu, India. His area of interests includes Medical Data Mining, Rough Set and Bioinformatics.